\documentstyle[preprint,aps,pre,psfig]{revtex}
\tightenlines
\newcommand{\be}{\begin{equation}}
\newcommand{\ee}{\end{equation}}
\newcommand{\ba}{\begin{eqnarray}}
\newcommand{\ea}{\end{eqnarray}}

\newcommand{\bpsi}{\mbox{\boldmath $\psi$}}
\newcommand{\bnabla}{\mbox{\boldmath $\nabla$}}
\newcommand{\bjmath}{\mbox{\boldmath $\jmath$}}
\newcommand{\bPi}{\mbox{\boldmath $\Pi$}}
\newcommand{\bpi}{\mbox{\boldmath $\pi$}}
\newcommand{\br}{{\bf r}}
\newcommand{\bk}{{\bf k}}

\newcommand{\eps}{\epsilon}

\begin{document}
\title{Spatial Correlations in Compressible Granular Flows}
\author{
T.P.C.~van Noije and M.H.~Ernst\\
{\it Instituut voor Theoretische Fysica, Universiteit Utrecht,
Postbus 80006,
3508 TA Utrecht, The Netherlands}
\\
\vspace*{5pt}
R.~Brito\\
{\it Facultad de Ciencias F\'{\i}sicas, Universidad Complutense,
28040 Madrid, Spain}}
\date{\today}
\maketitle
\begin{abstract}
For a freely evolving granular fluid, the buildup of spatial
correlations in density and flow field is described using
fluctuating hydrodynamics.
The theory for incompressible flows is
extended to the general, compressible case, including longitudinal
velocity and density fluctuations,
and yields qualitatively different results for long range
correlations.
The structure factor of density 
fluctuations shows a maximum at
finite wavenumber, shifting in time to smaller wavenumbers and 
corresponding
to a growing correlation length.
It agrees well 
with two-dimensional molecular dynamics simulations.
\end{abstract}
%\begin{multicols}{2}
\narrowtext
In most studies of {\em rapid granular flows}, also called the 
{\em granular
gas} regime \cite{jaeger}, the inelasticity of granular
collisions is assumed to be the most important feature that
distinguishes these flows from usual liquid or  
gas flows.
The dynamics is modeled by a single {\em inelasticity parameter} 
$\eps=1-\alpha^2$, where $\alpha$ is the coefficient of normal
restitution.
As a consequence a granular flow can only be maintained in
driven systems, where energy is put into the system e.g.\
by gravity, shear or in vibrated layers \cite{jaeger}.
Also quite some work has
been done on the freely evolving granular fluid 
\cite{goldhirsch,mcnamara,deltour,esipov,brey}, which has been
shown to be {\em linearly unstable} (onset of clustering
instability) with respect to spatial 
fluctuations in
density, $\delta n(\br,t)=n(\br,t)-\langle n \rangle$ \cite{goldhirsch}.
In Ref.\ \cite{noije}, an analytic description has been given of 
the buildup
of equal time spatial correlations in the flow field,
\be
G_{\alpha\beta}(\br,t)=\frac{1}{V} \int d{\bf r}^\prime \langle
u_\alpha({\bf r}+\br^\prime,t) u_\beta({\bf r}^\prime,t)\rangle,
\ee
of a system initialized in a spatially homogeneous state.
The theory is based on fluctuating hydrodynamics and
the assumption of incompressible flow, 
$\bnabla\cdot {\bf u}=0$.
This theory yields predictions, including long range tails
$\sim r^{-d}$, that  
for nearly elastic particles ($\eps\lesssim 0.2$)
agree well with 2-$D$ molecular dynamic simulations up to large distances.
Here we will extend the theory to the general,
compressible case, allowing us to calculate longitudinal velocity and 
density
fluctuations.
In fact, the structure factor corresponding to density
fluctuations, $S_{nn}(k,t)=V^{-1}\langle \delta n(\bk,t) \delta
n(-\bk,t) \rangle$, has been analyzed before by Deltour and Barrat
\cite{deltour}.
The difference between compressible and incompressible flow is
best
appreciated in Fourier space, where velocity correlations are
described by the tensor $S_{\alpha\beta}(\bk,t)$.
Both $G_{\alpha\beta}(\br,t)$ and its Fourier transform
$S_{\alpha\beta}(\bk,t)=V^{-1}\langle u_\alpha(\bk,t)
u_\beta(-\bk,t)\rangle$ are isotropic tensors and can be
decomposed into two scalar isotropic functions in the following
way:
\ba
G_{\alpha\beta}(\br,t)&=&\hat{r}_\alpha\hat{r}_\beta 
G_\parallel(r,t)+(\delta_{\alpha\beta}-\hat{r}_\alpha 
\hat{r}_\beta) 
G_\perp(r,t)\nonumber\\
S_{\alpha\beta}({\bf k},t)&=&\hat{k}_\alpha\hat{k}_\beta
S_\parallel(k,t)+(\delta_{\alpha\beta}-\hat{k}_\alpha\hat{k}_\beta)
S_\perp(k,t),
\ea
where carets denote unit vectors.
In a system of {\em elastic} hard spheres (EHS) for
times larger than the mean free time $t_0$, the correlation
functions are given by the equilibrium values, i.e.\
$G_{\alpha\beta}(\br,t)=[T/m n]\delta_{\alpha\beta}\delta(\br)$,
containing self-correlations only, and
$G_{nn}(r,t)= n\delta(\br) + n^2 [g(r)-1]$, where $g(r)$ is the
pair distribution function in thermal equilibrium.
For convenience we substract self-correlations and 
introduce the functions
$G^+_{\alpha\beta}(\br,t)\equiv G_{\alpha\beta}(\br,t)-
[T(t)/m n]\delta_{\alpha\beta} \delta(\br)$ and 
$S^+_{\alpha\beta}(\bk,t)\equiv
S_{\alpha\beta}(\bk,t)-[T(t)/m n]
\delta_{\alpha\beta}$.
Note that $T(t)$ is measured in energy units ($k_B=1$).
The structure factor of transverse velocity fluctuations,
$S_\perp^+(k,t)$, was calculated analytically in Ref.\ \cite{noije}
and shown to yield a long range $r^{-d}$ tail
in $G_\perp(r,t)$ and $G_\parallel(r,t)$ in case 
the fluctuations in the flow field are incompressible,
i.e. $S^+_\parallel(k,t)=0$.

In this Letter the structure factors
$S_{\alpha\beta}(\bk,t)$ and $S_{nn}(k,t)$ 
and corresponding spatial correlation
functions $G_{\alpha\beta}(\br,t)$ and $G_{nn}(r,t)$ will be
calculated and compared with 2-$D$ molecular dynamics simulations
for inelastic hard disk systems.
We show in particular by explicit calculation that for small
inelasticity ($\eps\lesssim 0.2$)
$S_\parallel^+(k,t)$ is essentially vanishing for all wavenumbers
except at very small $k$ values ($k\lesssim 1/\xi_\parallel$),
where the assumption of incompressible
$\bf u$ fluctuations, made in Ref.\ \cite{noije}, breaks down. 
Consequently, the most important role of $S_\parallel^+(k,t)$
is to provide an exponential cutoff for the $r^{-d}$ tail at the
largest scales $r \gtrsim 2\pi \xi_\parallel$.
At larger inelasticities the contributions from $S^+_\parallel(k,t)$
modify $G_\parallel(r,t)$ and $G_\perp(r,t)$ significantly at all
distances. 

The hydrodynamic equations for the 
unforced inelastic hard sphere (IHS) fluid
possess an exact solution, the {\em
homogeneous cooling state} (HCS), with a homogeneous density $n$
and temperature $T(t)$, and vanishing flow field.
Energy is dissipated at a rate 
$2 \gamma_0 \omega T$, where $\gamma_0=\eps/2d$ 
and where the collision frequency $\omega=\Omega_d \chi(n) n \sigma^{d-1}
\sqrt{T/\pi m}$ is calculated from the
Enskog-Boltzmann equation \cite{chapman} for a dense system of hard 
disks or
spheres ($d=2,3$).
Here $\Omega_d=2 \pi^{d/2}/\Gamma(d/2)$ is the
surface area of a $d$-dimensional unit sphere, $\sigma$ and $m$ the sphere
diameter and mass, both of which are set equal to unity, 
and $\chi(n)$ the pair 
correlation function at contact.
For detailed definitions and derivations we refer to
\cite{goldhirsch}.
In the following, we will assume that IHS hydrodynamics 
can be described by the standard hydrodynamic
equations supplemented by 
an energy sink term, which is evaluated in the local
homogeneous cooling state.
The equations of change for the macroscopic fields then become
\ba
&&\partial_t n +  \bnabla\cdot (n {\bf
u})=0\nonumber\\
&&\partial_t {\bf u}+{\bf u}
\cdot
\bnabla {\bf u}= -\frac{1}{n}  \bnabla\cdot  \bPi\\
&&\partial_t T +{\bf u}\cdot  \bnabla T
= -\frac{2}{d n}  (\bnabla\cdot {\bf J}
+\bPi: \bnabla {\bf u}) 
- 2 \gamma_0
\omega[n,T] T\nonumber,
\ea
where all fields and fluxes depend on $(\br,t)$.
A possible justification of these equations to lowest order in
$\eps$ can be found in Ref.\ \cite{goldhirsch2}, as well as a
discussion of higher order terms.
The pressure tensor $\bPi=\bPi_0 + \bPi_1$ is
given by 
$\bPi_0= p {\bf I}=n T (1+\Omega_d\chi n \sigma^d /2 d){\bf
I}$, with ${\bf I}$ the identity matrix, and
$\bPi_1= -2 \eta \{\bnabla {\bf u}\}_s 
 - \zeta
(\bnabla\cdot{\bf u}) {\bf I}$,
with $\{ \bnabla {\bf
u}\}_{s,\alpha\beta}=\textstyle{\frac{1}{2}}
(\nabla_\alpha u_\beta+\nabla_\beta u_\alpha-2\delta_{\alpha\beta}
\bnabla\cdot {\bf u}/d)$, and
the heat flow ${\bf J}=-\kappa \bnabla T$.
Here $\eta$, $\zeta$ and $\kappa$ are, respectively, the shear
viscosity, bulk viscosity and heat conductivity, given by the
Enskog theory for EHS with temperature $T(t)$
still 
depending explicitly on time to account for the homogeneous cooling
\cite{noije}.
The equations of change for the mesoscopic fields are obtained from the
above equations by adding fluctuating terms to the
pressure tensor and heat flow, denoted by $\hat{\bPi}$
and $\hat{\bf J}$ respectively \cite{landau}. They are characterized by a
vanishing average and correlations
which are local in space and time,
the strength of which is assumed to be determined by the standard
fluctuation-dissipation theorem:
\ba
&&\langle\hat{\Pi}_{\alpha\beta}(\br,t)\hat{\Pi}_{\gamma\delta}(\br^\prime,t^\prime)\rangle=2 T[\eta(\delta_{\alpha\gamma}\delta_{\beta\delta}
+\delta_{\alpha\delta}\delta_{\beta\gamma})\nonumber\\
&&\;\;\;\;\;\;\;\;\;\;\;\;\;\;\;\;\;\;
+(\zeta-\frac{2}{d}\eta)\delta_{\alpha\beta}\delta_{\gamma\delta}]\delta(\br-\br^\prime)\delta(t-t^\prime)\nonumber\\
&&\langle \hat{J}_\alpha
(\br,t)\hat{J}_\beta(\br^\prime,t^\prime)\rangle=2\kappa T^2
\delta_{\alpha\beta} \delta(\br-\br^\prime)\delta(t-t^\prime),
\ea
with transport coefficients depending on $T(t)$.
We are interested in the buildup of correlations between spatial
fluctuations in a system
that is prepared in a homogeneous state at an initial temperature
$T_0$ and reaches the HCS
within a few mean free times $t_0=1/\omega[n,T_0]$.
Therefore, we can linearize the 
above equations around a homogeneous density $n$ and a temperature
$T(t)=T_0 /[1+\gamma_0 t/t_0]^2$, and a vanishing flow field.
At this point it is convenient to make the change of variables,
$d\tau=\omega[n,T(t)] dt$, where $\tau$ is the average number of
collisions a particle has suffered within a time $t$,
$\delta n(\br,t)= n \delta \nu(\br,\tau)$, ${\bf u}(\br,t)=\sqrt{T(t)} {\bf w}(\br,\tau)$, 
$\delta
T(\br,t)=T(t) \delta \theta(\br,\tau)$, $\hat{\bPi}(\br,t)=n \omega[n,T(t)]
\sqrt{T(t)} \hat{\bpi}(\br,\tau)$ and $\hat{\bf J}(\br,t)=n \omega[n,T(t)]
T(t) \hat{\bjmath}(\br,\tau)$.
In these new variables the noise strengths of the reduced fluctuating 
pressure 
tensor $\hat{\bpi}$ and heat
flow $\hat{\bjmath}$ are {\em time independent},
and the equations of change for the mesoscopic Fourier modes $\delta
\nu(\bk,\tau)$, ${\bf w}(\bk,\tau)$ and $\delta \theta(\bk,\tau)$ become 
ordinary
differential equations with {\em time independent} coefficients (valid
for $k l_0\lesssim 1$ where $l_0=\sqrt{2 T(t)}/\omega[n,T(t)]$ is the
time independent mean free path): 
\ba
\frac{\partial \delta \nu}{\partial \tau} &=& - \frac{i k l_0}{\sqrt{2}}
w_l\nonumber\\
\frac{\partial w_{\perp\alpha}}{\partial \tau} &=& \gamma_0 (1-k^2
\xi_\perp^2)
w_{\perp\alpha}
- i k \hat{\pi}_{\alpha l}\nonumber\\
\frac{\partial w_l}{\partial \tau}&=&\gamma_0 (1-k^2 \xi_l^2) w_l
-\frac{i k l_0}{\sqrt{2}} \left(\frac{p}{n T}\right) 
\delta \theta\nonumber\\
&&- \frac{i k l_0}{\sqrt{2}} \left(\frac{1}{n T \chi_T}
\right)
\delta \nu
- i k \hat{\pi}_{ll}\nonumber\\
\frac{\partial \delta \theta}{\partial \tau}&=& -\gamma_0 (1+k^2
\xi_T^2) \delta \theta - \frac{i k l_0}{\sqrt{2}} \left(\frac{2 p}
{ d n
T}\right) w_l\nonumber\\
&&-2 \gamma_0 (1+\frac{n}{\chi}\frac{\partial
\chi}{\partial n})\delta \nu
- i k \frac{2}{d} \hat{\jmath_l}.
\label{eq:change}
\ea
Here we have introduced the time independent correlation lengths 
$\xi_\perp$, $\xi_l$ and
$\xi_T$, defined by $\xi_\perp^2=\nu/\omega
\gamma_0$ with $\nu=\eta/m n$ the kinematic viscosity, 
$\xi_l^2 =[2\nu (d-1)/d+\zeta/m n]/\omega\gamma_0$
and $\xi_T^2=2 \kappa/d n \omega \gamma_0$, and the isothermal
compressibility $\chi_T= {(\partial
n/\partial p)}_T/n$.
The subscript $\alpha$ in the equation for ${\bf w}_\perp$ refers
to any of the $(d-1)$ directions perpendicular to $\bk$, and the
subscript $l$ denotes the longitudinal direction along $\bk$.
To calculate the structure factors we also need the Fourier modes
with $\bk$ replaced by $-\bk$.

Since the transverse velocity ${\bf w}_\perp$ is
decoupled from the other modes, its structure factor $S_\perp(k,t)=\langle
u_{\perp\alpha}(\bk,t) u_{\perp\alpha}(-\bk,t) \rangle/V$ 
can be obtained
in the analytic form \cite{noije} 
\be
S_\perp(k,t)=
\frac{T(t)}{n}\left\{1+\frac{\exp[2\gamma_0\tau(1-k^2\xi_\perp^2)]-1}
{1-k^2\xi_\perp^2}\right\},
\label{eq:sperp}
\ee
which is valid for $k l_0\lesssim 1$.
The same result has been obtained from a more microscopic approach,
using {\em ring kinetic theory} \cite{noije2}.

The density, longitudinal velocity and temperature modes
are coupled and their equations of change can be written in matrix
representation as
\be
\frac{\partial}{\partial \tau}\bpsi(\bk) = {\bf M}(\bk)\bpsi(\bk) + 
\hat{\bf
f}(\bk),
\ee
where $\bpsi$ is the column vector
with components $\psi_1=\delta \nu$,
$\psi_2=w_l$ and $\psi_3=\delta \theta$,
and the hydrodynamic matrix ${\bf M}$ and the noise vector $\hat{\bf
f}$ are given by Eqs.\ (\ref{eq:change}).
Note that the elements $M_{31}(\bk)$ and $M_{33}(\bk)$, entering the
temperature equation, depend on the energy dissipation term.
In this notation the equal time correlations obey the equation of
change
\ba
&&\frac{\partial}{\partial \tau} \langle
\psi_\alpha(\bk,\tau)\psi_\beta(-
\bk,\tau)\rangle= M_{\alpha\gamma}(\bk)\langle
\psi_\gamma(\bk,\tau)\psi_\beta(-\bk,\tau)\rangle\nonumber\\
&&+M_{\beta\gamma}(-\bk) \langle \psi_\alpha(\bk,\tau)
\psi_\gamma(-\bk,\tau)\rangle + C_{\alpha\beta}(k),
\ea
where $\alpha,\beta,\dots=1,2,3$ label the components $\delta \nu$,
$w_l$ and $\delta \theta$.
These equations constitute a set of $3\times 3$ linear ordinary
differential equations, 
of which only 6 are independent.
The matrix of noise strengths $C_{\alpha\beta}(k)$, 
defined through $\langle
\hat{f}_\alpha(\bk,\tau)\hat{f}_
\beta(-\bk,\tau^\prime)\rangle= C_{\alpha\beta}(k)
\delta(\tau-\tau^\prime)$, has only two nonvanishing components, namely
$C_{22}=
2 V \gamma_0 k^2 \xi_l^2/n$ and
$C_{33}= 4 V
\gamma_0 k^2 \xi_T^2/d n$.
We have solved the above set of equations numerically, starting
from initial equilibrium correlations, 
of which the only nonvanishing ones are 
$\langle \psi_1(\bk,0) \psi_1(-\bk,0)\rangle=
V T \chi_T$,
$\langle \psi_2(\bk,0) \psi_2(-\bk,0)\rangle=V/n$
and $\langle \psi_3(\bk,0) \psi_3(-\bk,0)\rangle=
2 V/d n$ (for $k\ne 0$).
The most important new results with respect to Ref.\ \cite{noije} 
are the structure factors 
$S_\parallel(k,t)$ and $S_{nn}(k,t)$, and the 
correlation function 
$G_{nn}(r,t)$.
In Fig.\ 1, we show the results for these structure factors,
including
$S_\perp(k,t)$,
for a system with area fraction $\phi=0.245$ and $\alpha=0.9$ 
together with the results from a
molecular
dynamics simulation of 50000 inelastic hard disks.
We observe that $S_\parallel(k\rightarrow 0,t)=S_\perp(k\rightarrow
0,t)$,
implying for large distances an asymptotic behavior 
$G_{\alpha\beta}(\br,t)\sim S_\perp(k\rightarrow 0,t)
\delta_{\alpha\beta} \delta(\br)$, and thus the absence of
algebraic long range correlations on the largest scales ($r\gg 2\pi
\xi_\parallel$).
Therefore, we can already conclude that the asymptotic behavior of
$G_\perp(r,t)$ and $G_\parallel(r,t)$ cannot be $r^{-d}$.
Instead
the $r^{-d}$ tail obtained in Ref.\ \cite{noije} describes intermediate 
behavior which is exponentially cut off at
a distance determined by the width of $S^+_\parallel(k,t)$.
This width can be estimated from the eigenvalues of the
hydrodynamic matrix, more precisely from the dispersion relation of
the heat mode \cite{mcnamara}, 
which is a pure longitudinal velocity $w_l$ for $k\rightarrow 0$.
To second order in $k$ its dispersion relation 
is given by $z_H(k)=
\gamma_0(1-k^2 \xi_\parallel^2)$, with 
\be
\xi_\parallel^2=\xi_l^2
+\frac{l_0^2}{2\gamma_0^2}\left[\frac{1}{n T
\chi_T}-\frac{p}{n T}\left(1+\frac{n}{\chi}\frac{\partial
\chi}{\partial n} - \frac{p}{d n T}\right)\right].
\label{eq:heatapp}
\ee
Note that $\xi_\parallel\sim 1/\eps$ for small inelasticity,
whereas $\xi_\perp\sim \xi_l\sim \xi_T \sim 1/\sqrt{\eps}$
[see Fig.\ 2(c)].
To a good approximation $S_\parallel(k,t)$
for small wavenumbers is given by 
expression (\ref{eq:sperp}) with $\xi_\perp$ replaced
by $\xi_\parallel$.
This approximation is excellent up to wavenumbers where the exact
numerical result for $S_\parallel(k,t)$ shows a little dip 
(see Fig.\ 1, $\tau=19.4$,
$k\simeq 0.1$). 
At about the same wavenumber the structure factor $S_{nn}(k,t)$ 
reaches its maximal value, which grows in time.
The exact position of this maximum shifts in time to smaller
wavenumbers corresponding to a growing correlation length.
This can be explained by the following argument: for $k\rightarrow
0$ density
fluctuations $\delta n(\bk,t)$ are decoupled from the heat mode 
and we expect that
$S_{nn}(k\rightarrow 0,t)$ remains at its initial equilibrium value;
at small, but nonvanishing $k$, $\delta
n(\bk,t)$ couples in
${\cal O}(k)$ to the
unstable heat mode and the maximum of $k
\exp[2 z_H(k)\tau]$ shifts in time to smaller wavenumbers.

The estimate for $S_{nn}(k,t)\simeq S_{nn}(k,0) \exp[2 z_H(k)
\tau]$, used in Ref.\ \cite{deltour}, differs in two aspects from
our predictions:
(i) it neglects the wavenumber dependence of the coupling of density
fluctuations to the heat mode, giving for 
$S_{nn}(k,t)$ a decreasing function of $k$, 
and therefore cannot explain the
growing correlation length;
(ii) it neglects the fluctuating parts of the
pressure tensor and heat flow,
causing only numerical deviations from our prediction.
Note that seven out of the eight sets of data points shown 
in Fig.\ 9 of
Ref.\ \cite{deltour} are in the crossover or nonlinear time regime, 
which is 
estimated in Ref.\ \cite{noije} to occur at $\tau_{\rm cr}\simeq
65$ for $\alpha=0.9$ and $\phi=0.4$ and where our linear theory
breaks down.

Using the above approximation for $S_\parallel(k,t)$, 
the structure factor
$S^+_{\alpha\beta}(\bk,t)$ can be written as
\ba
S^+_{\alpha\beta}(\bk,t)&\approx&\frac{T(t)}{n}\int_0^s {\rm
d}s^\prime \exp(s^\prime)
\left[\hat{k}_\alpha\hat{k}_\beta \exp(-s^\prime
k^2\xi_\parallel^2) \right.
\nonumber\\
&&\left.+(\delta_{\alpha\beta}-\hat{k}_\alpha\hat{k}_\beta)\exp(-
s^\prime k^2 \xi_\perp^2)\right],
\ea
where $s=2\gamma_0 \tau$.
If the system is thermodynamically large ($L\gg 2\pi \xi_\parallel$),
$G^+_\parallel(r,t)=\hat{r}_\alpha\hat{r}_\beta
G^+_{\alpha\beta}(\br,t)$ and $G^+_\perp(r,t)=(\delta_{\alpha\beta}-
\hat{r}_\alpha
\hat{r}_\beta)
G^+_{\alpha\beta}(\br,t)/(d-1)$ can be obtained by performing 
integrals over $\bk$ space; the 
resulting $G^+_\parallel(r,t)$
and $G^+_\perp(r,t)$
can then be 
expressed as
integrals over simple functions. 
Here we only quote the results for $d=2$. Using 
\be
\int \frac{d{\bf q}}{(2\pi)^2} 
\sin^2{\theta} e^{i{\bf q}\cdot{\bf
x}-s q^2}=\frac{1}{2\pi x^2}[1-\exp(-x^2/4 s)],
\ee
where $\cos{\theta}=\hat{\bf q}\cdot\hat{\bf x}$,
we obtain
\ba
&&G^+_\lambda(r,t)\approx \frac{T(t)}{n}\left(\frac{1}{4\pi
\xi_\lambda^2} \int_0^s ds^\prime\frac{ 
\exp(s^\prime -x_\lambda^2/4 s^\prime)}{s^\prime}\right.\nonumber\\
&&\left.+\frac{m_\lambda}{2\pi r^2} \int_0^s d
s^\prime e^{s^\prime}\left[\exp\left(-\frac{x_\parallel^2}
{4 s^\prime}\right)-
\exp\left(-\frac{x_\perp^2}{4
s^\prime}\right)\right]\right)
\label{eq:expl}
\ea
for $\lambda=\parallel,\perp$, where $x_\lambda=r/\xi_\lambda$,
$m_\parallel=1$ and $m_\perp=-1$.
The approximation of {\em incompressible} fluid flow of Ref.\ 
\cite{noije} is obtained in the limit $\xi_\parallel\rightarrow
\infty$.
It is consistent with the thermodynamic concept of incompressibility,
i.e.\ $\chi_T=0$ in (\ref{eq:heatapp}), implying an infinite speed
of sound.
Fig.\ 2 shows $g_\parallel(x_\perp,s)=[n
\xi_\perp^2/T(t)] G^+_\parallel(r,t)$ and $g_\perp(x_\perp,s)=[n
\xi_\perp^2/T(t)] G^+_\perp(r,t)$ in the above
approximation for different
ratios
$\xi_\parallel/\xi_\perp$ at $s=2$.
In order to see $r^{-d}$ behavior the second term of 
(\ref{eq:expl}) should dominate over
the first.
For $\xi_\parallel^2 \gg \xi_\perp^2$, 
$\exp(-x_\perp^2/4 s^\prime)$ can be neglected with
respect to $\exp(-x^2_\parallel/4 s^\prime)$, and 
the second term
behaves algebraically if $x^2_\parallel\lesssim 4 s$. 
Restricting ourselves to $1\lesssim s\lesssim 10$ (i.e.\
moderate times where the correlations have
grown above noise level),
we can estimate that the range of algebraic decay is restricted to  
$r\lesssim 2\pi \xi_\parallel$,
where the factor $2\pi$ is chosen for convenience.
For $r\gtrsim 2\pi \xi_\parallel$, the remaining exponent cuts off the
$r^{-d}$ tail.

The predicted spatial velocity correlations 
$G_\parallel(r,t)$ and $G_\perp(r,t)$
have been obtained  
by numerically performing inverse Bessel
transformations on the numerical results for $S_\parallel(k,t)$ and
$S_\perp(k,t)$.
The result for $G_\parallel(r,t)$ corresponding to Fig.\ 1 
includes an intermediate
$r^{-2}$ tail and is shown in Fig.\ 3(a).
Fig.\ 3(b) shows the corresponding spatial density correlation
$G_{nn}(r,t)$ obtained numerically from $S_{nn}(k,t)$.
It confirms that the present theory correctly predicts
the buildup of density correlations, including a negative
correlation centered around a distance which grows in time as
$\sqrt{\tau}$.

At small inelasticity ($\eps \lesssim 0.2$) the functions 
$G_\parallel(r,t)$ and $G_\perp(r,t)$, calculated
here from the full set of hydrodynamic equations, differ for
$r\lesssim 2\pi \xi_\parallel$ only slightly from the results for
incompressible flow fields (see discussion in Ref.\ \cite{noije}).
However, the algebraic tails $\sim r^{-d}$ in $G_\parallel(r,t)$
and $G_\perp(r,t)$, derived in Ref.\ \cite{noije} for $r\gtrsim
2\pi \xi_\perp$, are exponentially cut off for $r\gtrsim
2\pi\xi_\parallel$.
As the correlation lengths $\xi_\perp\sim 1/\sqrt{\eps}$ and
$\xi_\parallel\sim 1/\eps$ are well separated for small
$\eps$ [see Fig.\ 2(c)], there is an intermediate range
of $r$ values where the algebraic tail $\sim r^{-d}$ in
$G_\parallel(r,t)$ can be observed.

At higher inelasticity $\xi_\parallel$ and $\xi_\perp$ are not well
separated and, as a consequence, there does not exist a spatial
regime in which the longitudinal fluctuations in the flow field can
be neglected and the regime of validity of the incompressible theory
of Ref.\ \cite{noije} has shrunk to zero.
Fig.\ 3(c) compares results from incompressible and compressible
fluctuating hydrodynamics with simulation data for $G_\perp(r,t)$ at
$\alpha=0.6$ and $\phi=0.4$ 
and confirms the validity of the 
compressible fluctuating hydrodynamics description up to reasonably 
large inelasticities.
Note that $G_{nn}(r,t)$ can only be calculated from the compressible
theory.

The authors want to thank J.A.G. Orza for performing simulations.
T.v.N. acknowledges support of the
foundation `Fundamenteel Onderzoek der Materie (FOM)', which is
financially supported by the Dutch National Science Foundation
(NWO).
R.B. acknowledges support from DGICYT (Spain) number PB94-0265.

%\end{multicols}
\begin{figure}
%\[ \psfig{file=Fig1.ps,width=8.5cm} \]
\caption{Theoretical predictions (solid lines) for $S_\perp(k,t)$, 
$S_\parallel(k,t)$ and $S_{nn}(k,t)$ versus $k\sigma$ for 
$\phi=0.245$ ($l_0\simeq
0.8$) and
$\alpha=0.9$, where
$\xi_\perp=4$ and $\xi_\parallel=17$, at $\tau=19.4$, 40 and 48.4,
compared with results from a single molecular 
dynamics run of $50000$ particles, implying a minimal
wavenumber $k_{\rm min}\sigma=2\pi\sigma/L \simeq 0.016$. All structure 
superimposed
on the plateau values
presents long range correlations of dynamic origin; 
equilibrium structure in $S_{nn}$ is only present for 
$k\sigma{\protect\gtrsim} 2\pi$.}
\end{figure}
\begin{figure}
%\[ \psfig{file=Fig2.ps,width=8.5cm} \]
\caption{(a) $^{10}\log{g_\parallel(x_\perp,s=2)}$ versus
$^{10}\log{x_\perp}$
from incompressible fluctuating hydrodynamics (solid line) and
present theory (\protect{\ref{eq:expl}}) (dashed lines from left to right:
$\xi_\parallel/\xi_\perp=1,2,5,10$);
as $\xi_\parallel/\xi_\perp$ decreases, the $r^{-2}$ tail
is cut off exponentially at smaller distances and
finally disappears at $\xi_\parallel=\xi_\perp$. (b)
$g_\perp(x_\perp,s=2)$ versus $x_\perp=r/\xi_\perp$;
the depth of the
minimum decreases with decreasing $\xi_\parallel/\xi_\perp$
and finally disappears at
$\xi_\parallel=\xi_\perp$.
(c) $\xi_\parallel$/$\xi_\perp$ versus area fraction $\phi$.}
\end{figure}
\begin{figure}[h]
%\[ \psfig{file=Fig3.ps,width=8.5cm} \]
\caption{(a) $^{10} \log[|G_\parallel|/T]$ versus $^{10}\log{r}$. 
(b) 
$G_{nn}$ versus $r$;
the same parameters $\alpha,\phi,\tau$ as in Fig.\ 1 are used for
(a) and (b).
(c) $G_\perp/T$ versus $r$ for $\phi=0.4$ ($l_0\simeq
0.34$), $\alpha=0.6$ ($\xi_\perp=1.46$, $\xi_\parallel=3.8$) at
$\tau=20$, 40 and 60; 
in (a) and (c) the solid (dashed) line is the
prediction from compressible (incompressible) fluctuating
hydrodynamics.}
\end{figure}
%\end{multicols}
\begin{center}
\pagebreak
 
\vspace*{5cm}
\thispagestyle{empty}
 
\begin{figure}
\[
\psfig{file=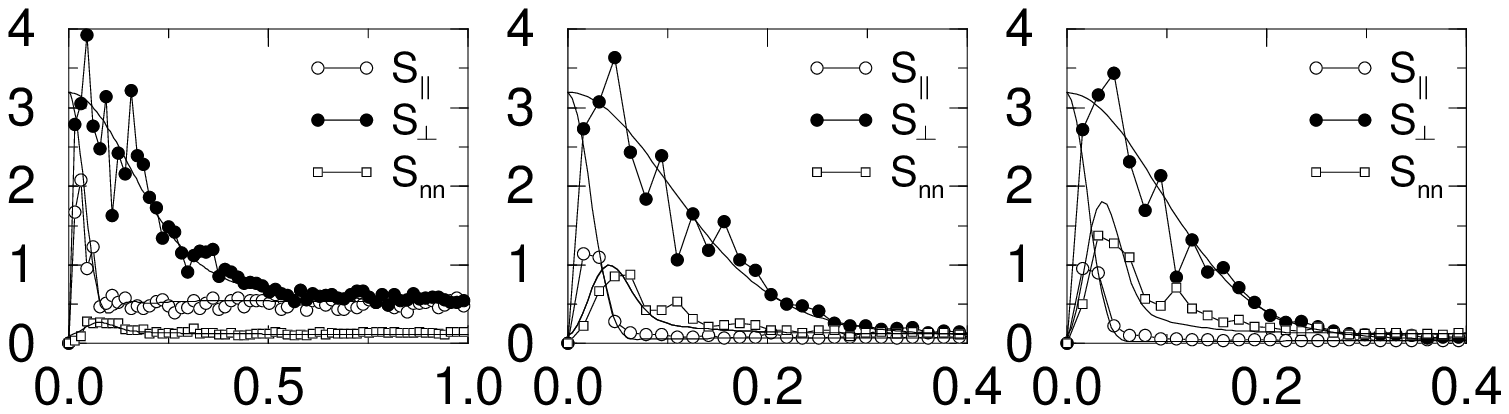}
\]
\end{figure}
 
\Large
 
Figure 1 \\[3mm]
 
\large
 
{\bf Title:} Spatial Correlations in Compressible Flows \\[2mm]
 
{\bf Authors:} T.P.C.~van Noije, M.H.~Ernst and R.~Brito
\newpage
 
\vspace*{5cm}
\thispagestyle{empty}
 
\begin{figure}
\[
\psfig{file=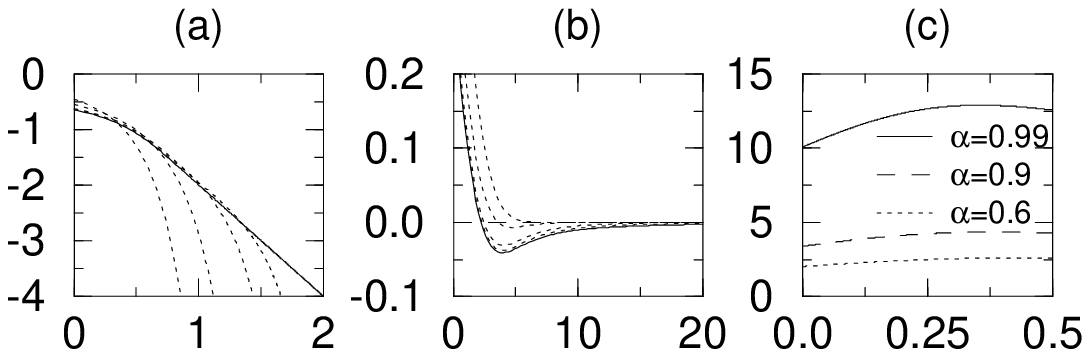}
\]
\end{figure}
 
\Large
 
Figure 2 \\[3mm]
 
\large
 
{\bf Title:} Spatial Correlations in Compressible Flows \\[2mm]
 
{\bf Authors:} T.P.C.~van Noije, M.H.~Ernst and R.~Brito 
\newpage
 
\vspace*{5cm}
\thispagestyle{empty}
 
\begin{figure}
\[
\psfig{file=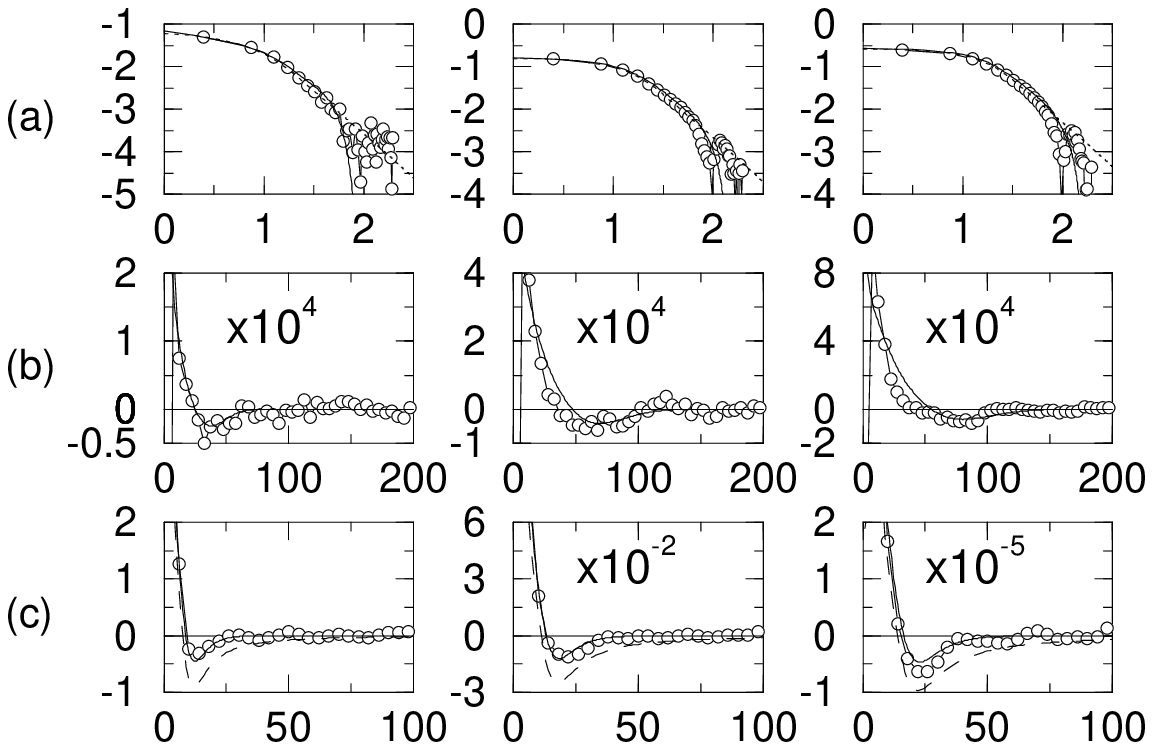}
\]
\end{figure}
 
\Large
 
Figure 3 \\[3mm]
 
\large
 
{\bf Title:} Spatial Correlations in Compressible Flows \\[2mm]
 
{\bf Authors:} T.P.C.~van Noije, M.H.~Ernst and R.~Brito 
\end{center}
\end{document}